\documentclass[%
 aip,
 amsmath,amssymb,
 reprint,%
]{revtex4-1}

\usepackage{graphicx}
\usepackage{subfigure}
\usepackage{dcolumn}
\usepackage{bm}

\usepackage[utf8]{inputenc}
\usepackage[T1]{fontenc}
\usepackage{mathptmx}
\usepackage{amssymb}
\usepackage{epstopdf}

\usepackage[breaklinks=true,
colorlinks=true,
linkcolor=blue,
citecolor=blue,
urlcolor=blue]{hyperref}

\begin{document}

\preprint{AIP/123-QED}

\title{Influences of sinusoidal density modulation on  stimulated Raman scattering in inhomogeneous plasmas}

\author{Y. Chen}
\affiliation{Key Laboratory for Micro-/Nano-Optoelectronic Devices of Ministry of Education, School of Physics and Electronics, Hunan University, Changsha, 410082, China}
\affiliation{Institute of Applied Physics and Computational Mathematics, Beijing, 100094, China}

\author{C. Y. Zheng}
\affiliation{Institute of Applied Physics and Computational Mathematics, Beijing, 100094, China}
\affiliation{\mbox{HEDPS, Center for Applied Physics and Technology, Peking University, Beijing 100871, China}}
\affiliation{\mbox{Collaborative Innovation Center of IFSA (CICIFSA), Shanghai Jiao Tong University, Shanghai 200240, China}}

\author{Z. J. Liu}
\affiliation{Institute of Applied Physics and Computational Mathematics, Beijing, 100094, China}
\affiliation{\mbox{HEDPS, Center for Applied Physics and Technology, Peking University, Beijing 100871, China}}

\author{L. H. Cao}
\affiliation{Institute of Applied Physics and Computational Mathematics, Beijing, 100094, China}
\affiliation{\mbox{HEDPS, Center for Applied Physics and Technology, Peking University, Beijing 100871, China}}
\affiliation{\mbox{Collaborative Innovation Center of IFSA (CICIFSA), Shanghai Jiao Tong University, Shanghai 200240, China}}

\author{Q. S. Feng}
\affiliation{Institute of Applied Physics and Computational Mathematics, Beijing, 100094, China}
\affiliation{\mbox{HEDPS, Center for Applied Physics and Technology, Peking University, Beijing 100871, China}}

\author{Y. G. Chen}
\affiliation{Key Laboratory for Micro-/Nano-Optoelectronic Devices of Ministry of Education, School of Physics and Electronics, Hunan University, Changsha, 410082, China}

\author{Z. M. Huang}
\affiliation{Key Laboratory for Micro-/Nano-Optoelectronic Devices of Ministry of Education, School of Physics and Electronics, Hunan University, Changsha, 410082, China}

\author{C. Z. Xiao } 
  \email{xiaocz@hnu.edu.cn}
  \affiliation{Key Laboratory for Micro-/Nano-Optoelectronic Devices of Ministry of Education, School of Physics and Electronics, Hunan University, Changsha, 410082, China}
  \affiliation{\mbox{Collaborative Innovation Center of IFSA (CICIFSA), Shanghai Jiao Tong University, Shanghai 200240, China}}
\date{\today}

  \begin{abstract}

     The influence of sinusoidal density modulation on the stimulated Raman scattering (SRS) reflectivity in inhomogeneous plasmas is studied by three-wave coupling equations, fully kinetic Vlasov simulations and particle in cell (PIC) simulations. Through the numerical solution of three-wave coupling equations, we find that the sinusoidal density modulation is capable of inducing absolute SRS even though the Rosenbluth gain is smaller than $\pi$, and we give a region of modulational wavelength and amplitude that the absolute SRS can be induced, which agrees with early studies.  The average reflectivity obtained by Vlasov simulations has the same trend with the growth rate of absolute SRS obtained by three-wave equations. Instead of causing absolute instability, modulational wavelength shorter than a basic gain length is able to suppress the inflation of SRS through harmonic waves. And, the PIC simulations qualitatively agree with our Vlasov simulations. Our results offer an alternative explanation of high reflectivity at underdense plasma in experiments, which is due to long-wavelength modulation, and a potential method to suppress SRS by using the short-wavelength modulation.

  \end{abstract}


  \maketitle

\section{Introduction}\label{introduction}
The stimulated Raman scattering (SRS) is an important physics process in laser plasma interaction (LPI), in which the pump laser decays to a reflected wave and a Langmuir wave. In inertial confinement fusion (ICF),\cite{ICF1,ICF2,ICF3,xiao2,xiao3,xiao4} this instability can take away a huge amount of energy from pump laser, which is detrimental to the fusion.
The physics of SRS in homogeneous plasma is well understood,\cite{kruer,Nicholson,xiao1,yanxia,haochu,chen} and it corresponds to  indirect-drive scheme.\cite{I_D}  On the other hand, the studies of SRS in inhomogeneous plasma are more common in other ICF schemes, such as direct-drive scheme,\cite{D_D} hybrid-drive scheme\cite{H_D} and shock ignition.\cite{S_I}
At early $70$s, Rosenbluth studied the parametric instabilities in  inhomogeneous plasma using WKB approximation and gave the well-known Rosenbluth gain.\cite{rosenbluth1,rosenbluth2} However, Rosenbluth's results are based on the ideal laser and plasma conditions. In subsequent studies, people found that some disturbance in laser or plasma would destabilize the Rosenbluth convective instability and turn it to absolute instability. Laval \emph{et al.} studied that the imperfect pump laser contributed to the development of absolute instabilities.\cite{laval} Picard and Johnston showed that the absolute instability could also be induced by  sufficiently strong pump laser.\cite{picard1}   Li \emph{et al.} showed that the long wave sinusoidal density modulation could induce the absolute two-plasmon-decay(TPD) and SRS near $n_{c}/4$  surface through theory and fluid simulations,\cite{lij} where $n_{c}$ is the critical density for pump laser.

More importantly, Picard and Johnston studied that the sinusoidal density modulation could induce the absolute instability and gave two kinds of threshold of absolute instability, one is the exact numerical solution and the other  is a quadratic fit formula.\cite{picard2} In Picard's exact numerical solution, the threshold of absolute instability has a clear cut off at the short modulational wavelength region, however the quadratic fit formula does not have this cut off, and we should note that the quadratic fit formula agree with exact solution at long wavelength region.  And, Nicholson  pointed out that this kind of absolute instability is sensitive to the wavelength of sinusoidal density modulation.\cite{sin} The growth rate of absolute instability falls off rapidly when the modulational wavelength  far away from $2\pi L_{0}$, where $L_{0}$  is the basic gain length for the absolute instability in a finite-length system (defined fully later).
Actually, we have found by simulations that modulational wavelength should be at least higher than $1.3$$L_0$ to trigger absolute instability. This finding explains Nicholson's simulation results\cite{sin} and is consistent with  Picard's exact numerical solution \cite{picard2} and  other studies that uses imperfect laser and plasma conditions.\cite{laval,picard1,lij,sin,turbulent,white,reiman} Besides, earlier researches only considered the influence of density modulation on instability at linear stage, the physics at nonlinear stage is still unknown.

In this paper, firstly, through the numerical solution of three-wave coupling equations, we find that the reflectivity of SRS will show an absolute feature  and the threshold of absolute SRS is agreed with Picard's theoretical formula\cite{picard2} when the modulational wavelength is close to $2\pi L_{0}$. However, the absolute SRS disappears when the modulational wavelength is less than $ 1.3 L_{0}$, which agrees well with Picard's exact solution. And we obtain a region of modulational wavelength and amplitude that density modulation can cause the absolute SRS.
Secondly, we have carried out the Vlasov simulations\cite{feng,wangqing2,yang} to study the influence of density modulation at linear and nonlinear stage. At linear stage, the Vlasov simulations show good agreement with three-wave equations analysis. The reflectivity of SRS will increase greatly when the absolute instability is induced and the short-wavelength density modulation has no influence on the reflectivity. Whereas, at nonlinear stage, especially when the inflation of SRS is induced by particle trapping,\cite{hxvu1,hxvu2,hxvu3} the short-wavelength modulation shows significant effect on suppressing the inflation of SRS, because the short-wavelength modulation increase the landau damping of Langmuir wave, and the downward harmonic wave can take away the energy of Langmuir wave. Thus the short-wavelength sinusoidal density modulation can be applied in the high intensity and inhomogeneous ignition schemes\cite{D_D,H_D,S_I,Klimo1,Klimo2,HaoL1} to suppress SRS. Finally, we use particle in cell (PIC) simulations to prove our conclusions.

This paper is structured in the following ways. Firstly, in Sec.~\ref{theoretical model}, we describe the three-wave coupling equations analysis of SRS in an inhomogeneous plasma. Secondly, Vlasov simulations and PIC simulations to verify our theoretical results are performed in Sec.~\ref{Simulation model} and Sec.~\ref{PIC model}. At last, the conclusion  and discussion about the sinusoidal density modulation are shown in  Sec.~\ref{conclusion}.

\section{Theoretical model}\label{theoretical model}

\subsection{ Numerical solution of three-wave coupling equations  }\label{twe}

The three-wave coupling equations in one dimension are frequently used to study  parametric instabilities in inhomogeneous plasmas, \cite{rosenbluth1,rosenbluth2,turbulent,sin}
\begin{equation} \label{wkb1}
    \begin{split}
    &(\partial_{t}+\nu_{0}+V_{0}\partial_{x})a_{0}(x,t)=-\gamma_{0} a_{1}(x,t)a_{2}(x,t),\\
    &(\partial_{t}+\nu_{1}+V_{1}\partial_{x})a_{1}(x,t)=\gamma_{0} a_{2}^{*}(x,t)a_{0}(x,t),\\
    &(\partial_{t}+\nu_{2}+V_{2}\partial_{x} + iV_{2}\kappa)a_{2}(x,t)=\gamma_{0}a_{0}(x,t)a_{1}^{*}(x,t),
    \end{split}
   \end{equation}
   where $a_{0}(x,t)$, $a_{1}(x,t)$ and $a_{2}(x,t)$ are slowly varying  amplitudes of three waves: pump wave, back reflected wave and plasma wave, respectively. $\nu_{0}$, $\nu_{1}$ and $\nu_{2}$ are the corresponding damping rates, $V_{0}$, $V_{1}$ and $V_{2}$ are the group velocities of three waves. $\gamma_{0}$ is the temporal growth rate of the instability in a homogeneous plasma without damping, and  $\kappa$ is the wave number mismatch of parametric instabilities in the inhomogeneous plasma.

   When a sinusoidal density modulation is added into the system, the plasma density profile becomes $n_{e}(x) = n_{e0}[1+(x-x_{r})/L + \varepsilon sin(k_{s}x)]$ which, in Eq.~(\ref{wkb1}), is depicted by the wave number mismatch $\kappa=\kappa'\cdot(x-x_{r})+\kappa_{m}sin(k_{s}x)$,  where $L$ is the density scale length, $\kappa' \approx \omega_{pe}^{2} / 6Lv_{e}^{2}k_{a_{2}}$ is the coefficient of spatial wave number detuning, $\kappa_{m}sin(k_{s}x)$ is the wave number mismatch caused by density modulation, and $x_{r}$ is the phase matching point for the instability. In the modulation, $k_{s} \equiv 1/L_{m}$ indicates modulational wave number, corresponding to a wavelength of $\lambda_{s} = 2 \pi L_{m}$, and  $\varepsilon = \delta n / n_{e0}$ is the modulational amplitude, which is related to the $\kappa_{m}$ through\cite{sin}
   \begin{equation} \label{wkb1_5}
     \kappa_{m}=\frac{k_{a_{2}}\varepsilon}{6(k_{a_{2}}\lambda_{De})^{2}}  ,
  \end{equation}
  where $k_{a_{2}}$ is the wave number of plasma wave, $\lambda_{De} = v_{e}/\omega_{pe}$ is the Debye length of electron at phase matching point, $v_{e}$ is the electron thermal velocity, and $\omega_{pe}= \sqrt{4\pi n_{e}e^{2} / m_{e}}$ is the electron plasma frequency.
   Without density modulation,  Eq.~(\ref{wkb1})  has analytic solutions,\cite{rosenbluth2} however, there is no  analytic solution when a sinusoidal density modulation has been taken into consideration.

   In order to obtain numerical solutions, we have developed a code numerically solving Eq.~(\ref{wkb1}) by Lax-Wendroff scheme,\cite{sweby,wangqing1} and we call it three-wave simulation below.  For simplicity, the damping rates of three waves are ignored in our three-wave simulations.  We assume that the SRS occurs at $n_{e0}=0.1n_{c}$, $L =$ $100 \mu$m,  and the temperature of electron is $T_{e}=1keV$. Then, the frequencies and group velocities of three waves can be calculated by dispersion relations,
\begin{equation} \label{wkb2}
 \begin{split}
   & \omega_{a_{0}}^{2} = \omega_{pe}^{2} + k_{a_{0}}^{2} c^{2},\\
   & \omega_{a_{1}}^{2} = \omega_{pe}^{2} + k_{a_{1}}^{2} c^{2},\\
   & \omega_{a_{2}}^{2} = \omega_{pe}^{2} + 3k_{a_{2}}^{2} v_{e}^{2},\\
   & \omega_{0}=\omega_{a_{1}}+\omega_{a_{2}}, k_{a_{0}} = k_{a_{1}}+k_{a_{2}}\\
   &  V_{0} = k_{a_{0}}c^{2}/ \omega_{a_{0}}, V_{1} = -k_{a_{1}}c^{2}/ \omega_{a_{1}}, V_{2} = 3k_{a_{2}}v_{e}^{2}/ \omega_{a_{2}}
 \end{split}
\end{equation}
  These give $\omega_{a_{0}}=\omega_{0}$, $V_{0}=0.9487c$, $\omega_{a_{1}}=0.66 \omega_{0}$, $V_{1}=-0.8787c$, $\omega_{a_{2}}=0.34 \omega_{0}$ and $V_{2}=0.0267c$, where $\omega_{0}$ is the frequency of pump laser, whose wavelength is $\lambda_{0} = 0.351 \mu$m, and $c$ is the speed of light in vacuum. $k_{a_{0}}$ , $k_{a_{1}}$ and $k_{a_{2}}$ are the wave numbers of pump laser, reflected laser and Langmuir wave, respectively.

  In the three-wave simulations, we split space and time with $\Delta x = 0.05 c/\omega_{0}$ and $\Delta t = 0.05 \omega_{0}^{-1}$.  The simulation box is $x = [-2000 , 2000 ] c/\omega_{0}$, the resonant point, $n_{e} = 0.1 n_{c}$, is located at $x_{r} = 0 $, and the simulation time is $t = 20000 \omega_{0}^{-1} = 3.5 ps$. The intensity of pump laser is $4 \times 10^{15} W/cm^{2}$, and then the growth rate of SRS can be calculated by $ \gamma_{0} =  \frac{k_{a_{2}}v_{os}}{4} \left[  \frac{\omega_{pe}^{2}}{\omega_{a_{2}}(\omega_{0}-\omega_{a_{2}})} \right]^{1/2} = 0.00486 \omega_{0}$, where $v_{os}$ is the oscillation velocity of electrons in the pump laser.  Initial conditions of three waves are $a_{0} (x, t=0) = 1 $, $a_{1} (x, t=0) = 1\times10^{-4}$ and $a_{2} (x, t=0) = 0$.  Based on the Rosenbluth's analytical results,\cite{rosenbluth2} the maximum reflectivity of convective SRS in an inhomogeneous plasma with linear density profile is
\begin{equation} \label{wkb3}
R = a_{1}^{2} exp(2G_{R}).
\end{equation}
In our conditions, the Rosenbluth gain is $G_{R}= \frac{\pi\gamma_{0}^{2}}{\kappa^{'}|V_{1}V_{2}|} = 1.021$.

\subsection{Three-wave simulation results}\label{twe1}

\begin{figure}
    \begin{minipage}[t]{0.5\linewidth}
    \centering
    \includegraphics[width=1.96in]{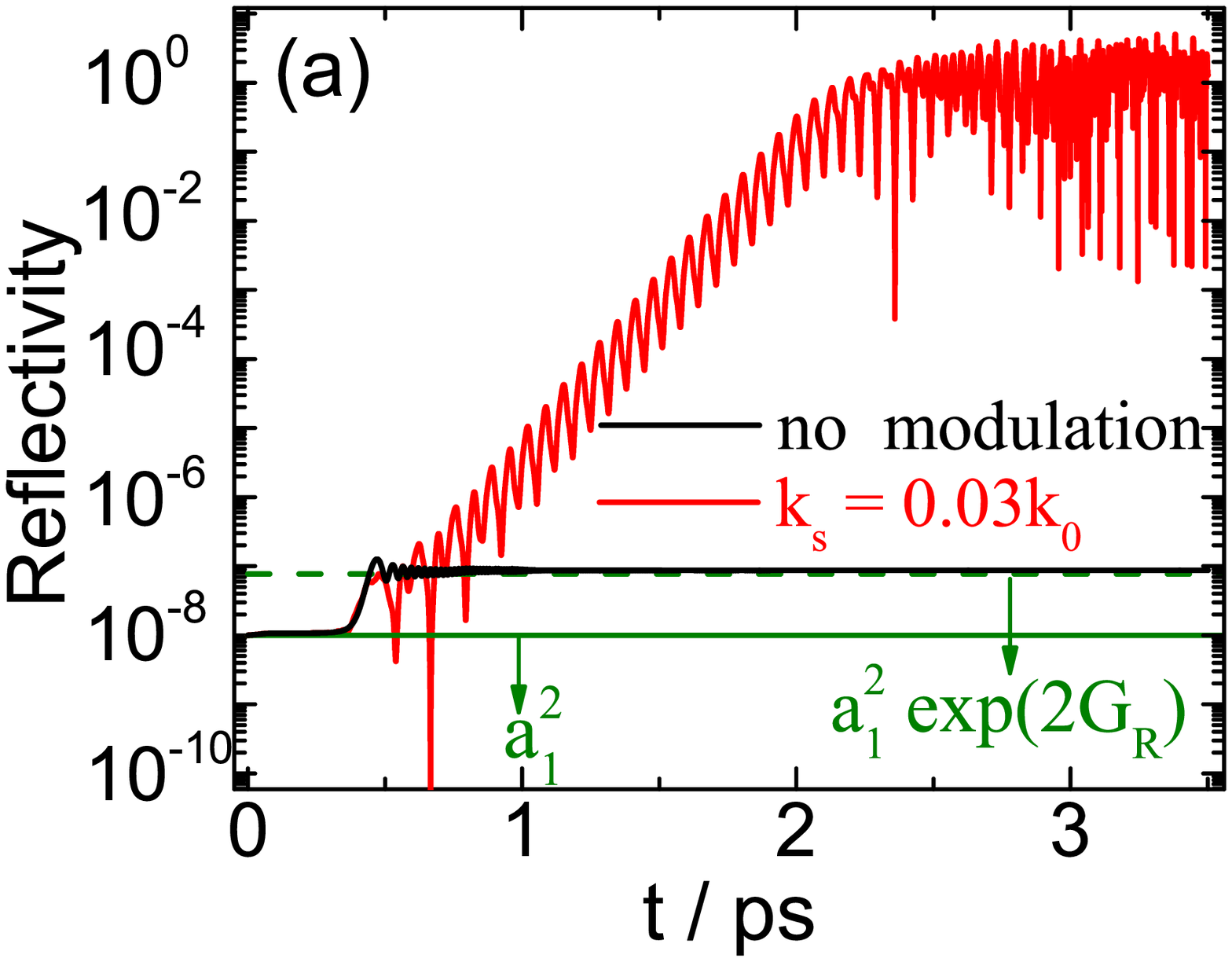}\vspace{-18pt}
    \label{fig:side:a}
    \end{minipage}%
    \begin{minipage}[t]{0.5\linewidth}
    \centering
    \includegraphics[width=1.96in]{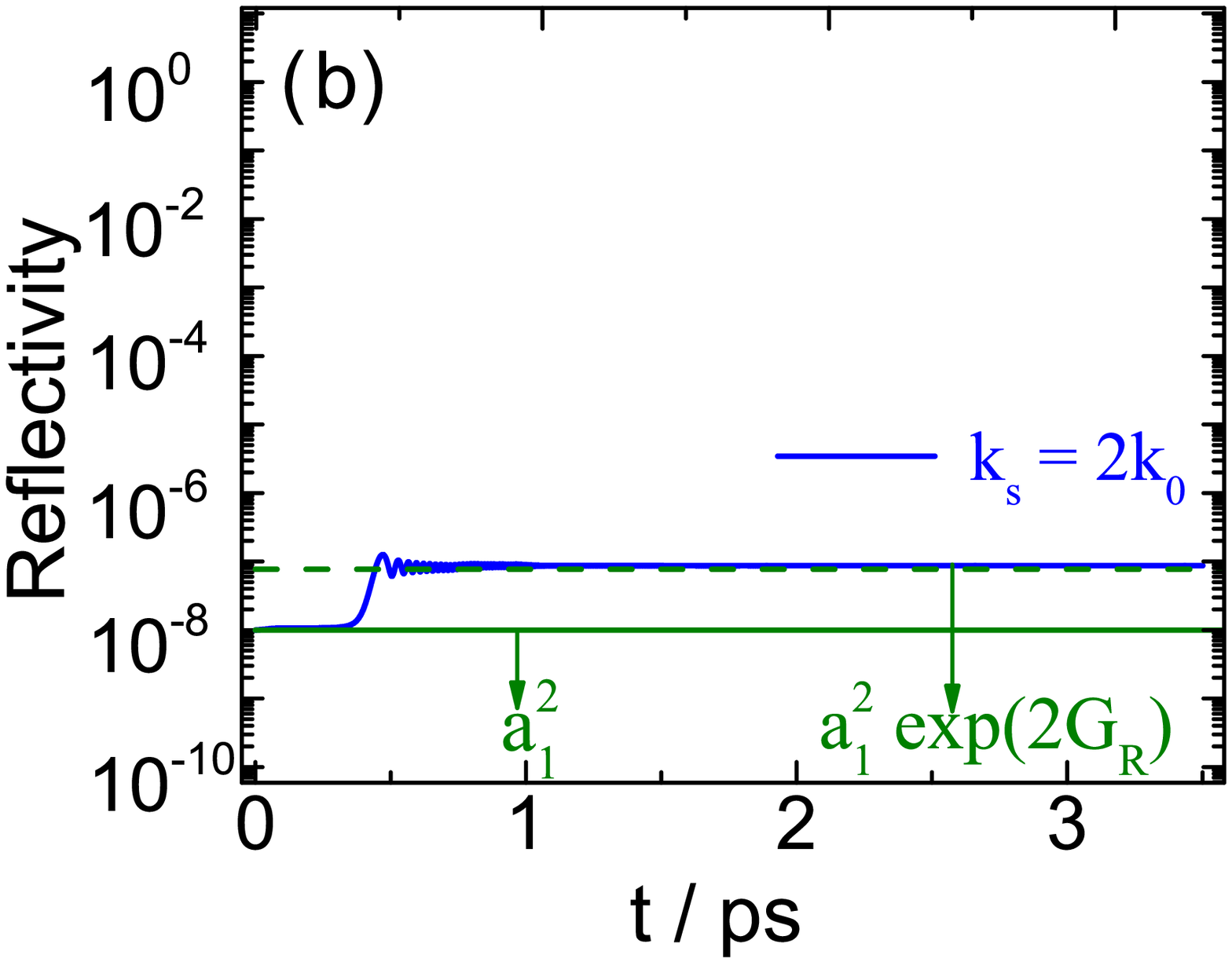}\vspace{-18pt}
    \label{fig:side:b}
    \end{minipage}
    \caption{\label{three_wave}(a) The Rosenbluth reflectivity in inhomogeneous plasma(green dashed line), the reflectivity that is obtained by three-wave simulation without density modulation(black line), and the reflectivity that is obtained by three-wave simulation with density modulation $k_{s} = 0.03k_{0}$, $\kappa_{m}=0.1 \omega_{0} / c$ (red line). (b) The reflectivity that is obtained by three-wave simulation with density modulation $k_{s} = 2k_{0}$, $\kappa_{m}=0.1 \omega_{0} / c$ (blue line). }
\end{figure}

   Fig.~\ref{three_wave} shows our three-wave simulation results, where the green solid line represents the initial seed of $a_{1}$, and the green dashed line stands for the reflectivity of this system obtained by Rosenbluth theory. First, in Fig.~\ref{three_wave}(a), we have carried on the three-wave simulation without density modulation represented by the black line, which agrees with the reflectivity calculated by the  Rosenbluth gain excellently.\cite{rosenbluth1,rosenbluth2}

  In Ref.~[\onlinecite{sin,picard2,lij}],  absolute instability will appear when sinusoidal density modulation exists in the system. There are two parameters relevant to the density modulation, the wave number $k_{s}$ and the amplitude $\kappa_{m}$, so the temporal growth rate of SRS is related to these  two parameters, and the absolute instability threshold given by Picard's\cite{picard2} theoretical formula is,
\begin{equation} \label{wkb4}
  \kappa_{m} > 2\frac{\kappa'}{k_{s}} exp(-2\frac{k_{s}}{\kappa'L_{0}} -1 ),
\end{equation}
once again, we note that this formula suits for long-wavelength modulation, the exact solution can be seen in Fig 10 of Ref.~[\onlinecite{picard2}]. Similar to earlier researches, we study the influence of density modulation on SRS. We consider two different kinds of density modulations, one is that the wavelength of density modulation is close to $2\pi L_{0}$, and another is that the wavelength of density modulation is much smaller than $2\pi L_{0}$, where $L_{0} = |V_{1}V_{2}|^{1/2}/\gamma_{0}= 33.3 c/\omega_{0}$.

First, we add a long-wavelength density modulation to the system. The modulational wave number is $k_{s} = 0.03 k_{0}$, $i.e.$, $L_{m}$ is equal to $L_{0}$, and the amplitude is $\kappa_{m} = 0.1 \omega_{0} / c$. The amplitude at real space is $\varepsilon = 0.018$, which is obtained by using Eq.~(\ref{wkb1_5}). The density modulation used here is  above the threshold of Picard's  formula\cite{picard2}, so the absolute SRS will be induced. As expected, in Fig.~\ref{three_wave}(a), the red line increases exponentially over time, which is a clear feature of absolute SRS , and when $t>2.1 ps$, the reflectivity becomes flat, it is because the pump depletion happens. Thus the average reflectivity of red line will be much larger than the average reflectivity obtained by Rosenbluth gain.

\begin{figure}
    \begin{minipage}[t]{0.5\linewidth}
    \centering
    \includegraphics[width=1.95in]{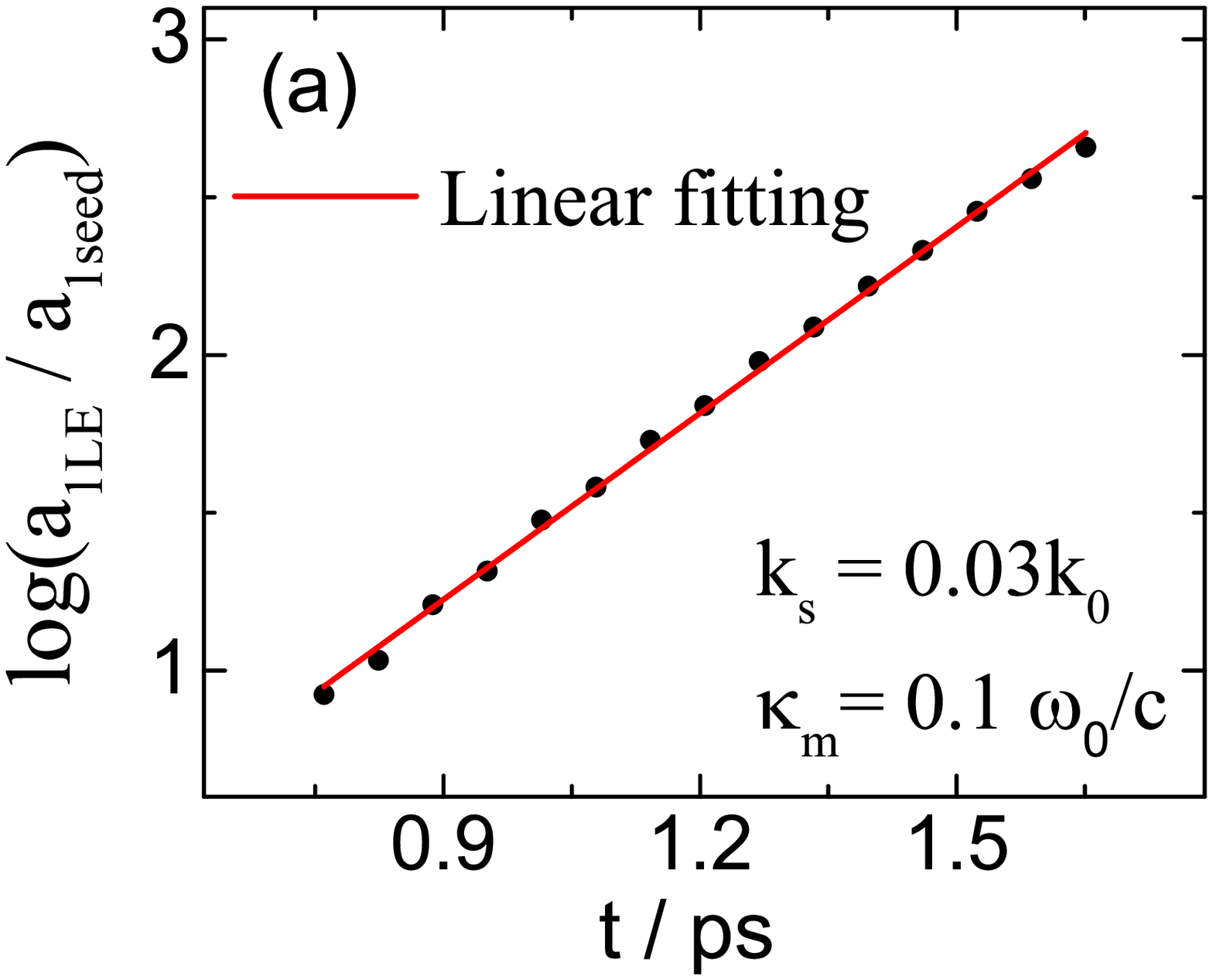}\vspace{-20pt}
    \label{fig:side:a}
    \end{minipage}%
    \begin{minipage}[t]{0.5\linewidth}
    \centering
    \includegraphics[width=1.95in]{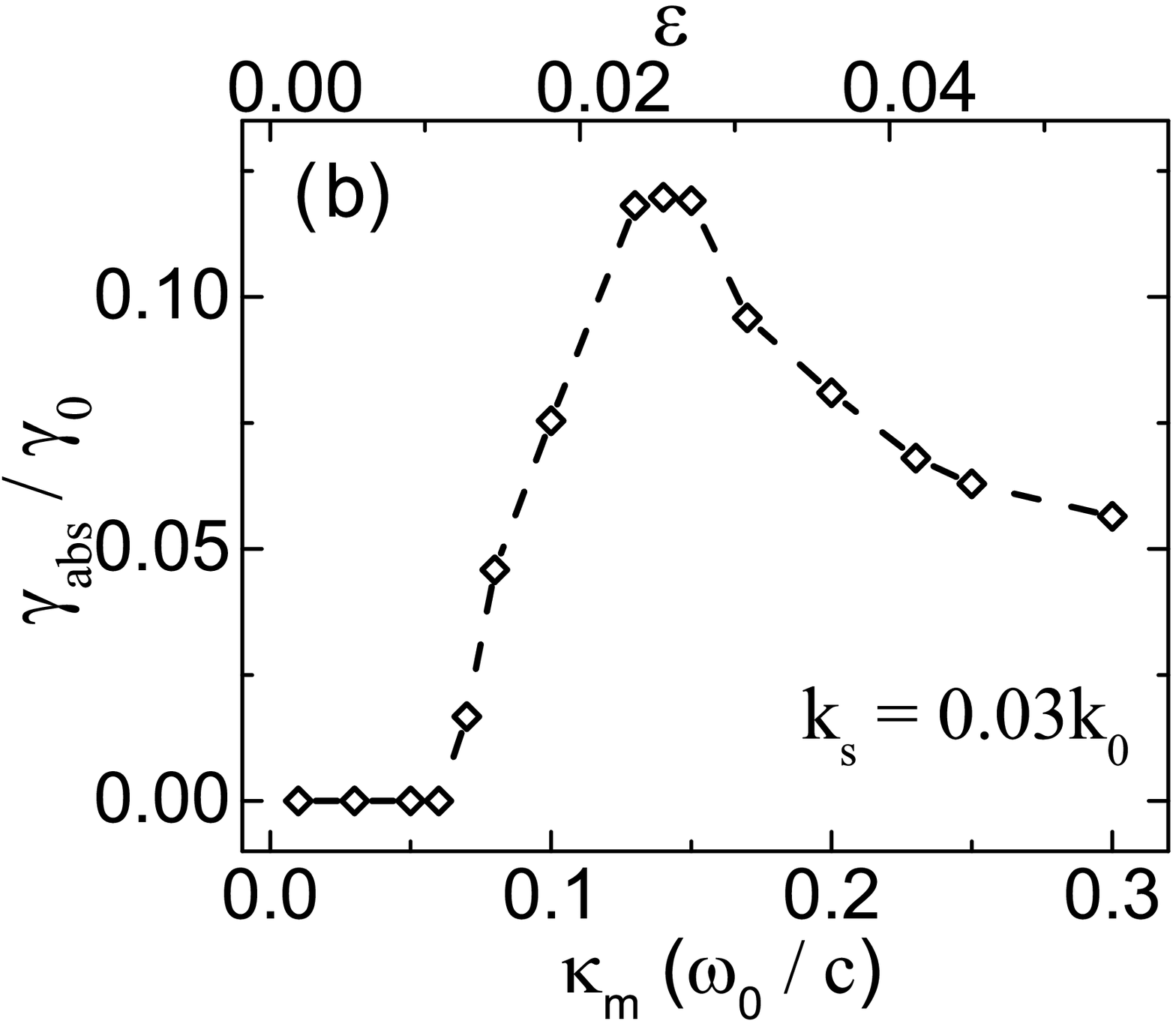}\vspace{-20pt}
    \label{fig:side:b}
    \end{minipage}
    \begin{minipage}[t]{0.5\linewidth}
    \centering
    \includegraphics[width=1.95in]{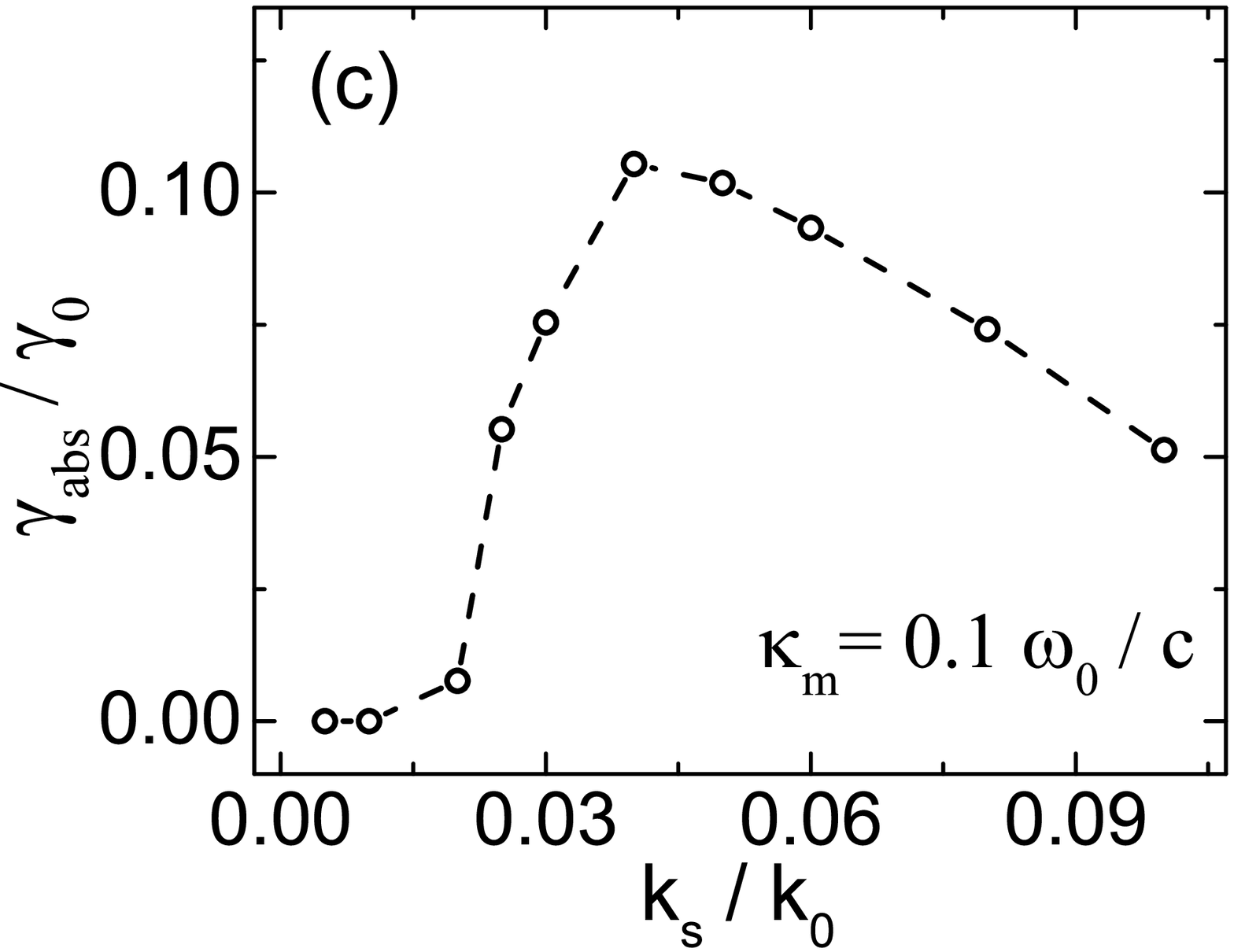}\vspace{-20pt}
    \label{fig:side:c}
    \end{minipage}%
    \begin{minipage}[t]{0.5\linewidth}
    \centering
    \includegraphics[width=1.95in]{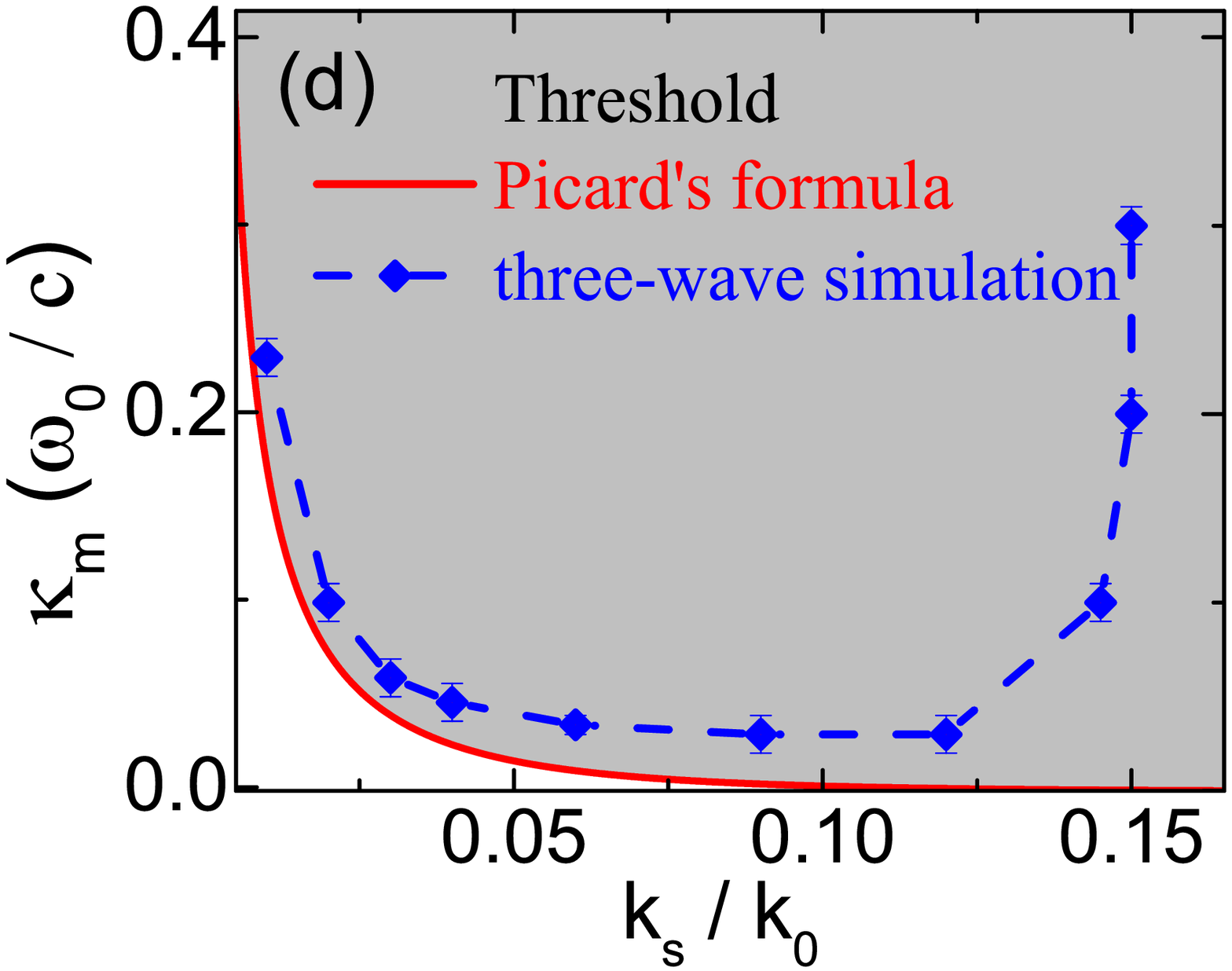}\vspace{-20pt}
    \label{fig:side:d}
    \end{minipage}

    \caption{\label{three_wave_2} (a) The linear fitting of $log(a_{1L} / a_{1seed})$ to obtain the growth rate of absolute SRS with density modulation $k_{s} = 0.03 k_{0}$, $\kappa_{m} = 0.1 \omega_{0} / c$. (b) The amplitude $\kappa_{m}$ dependence of growth rate with $k_{s} = 0.03 k_{0}$. (c) The wave number $k_{s}$ dependence of growth rate with $\kappa_{m} = 0.1 \omega_{0} / c$. (d) The threshold of absolute SRS at at two dimension plane of $k_{s}$ and $\kappa_{m}$ with Eq.~(\ref{wkb4}) (red line), the grey shaded area is the region that absolute SRS can occur. And the threshold obtained by three-wave simulations (blue diamonds).}
\end{figure}

Besides, the temporal growth rate of absolute SRS caused by density modulation can  be also obtained by three-wave simulations. As shown in Fig.~\ref{three_wave_2}(a), the time history of $a_{1}$  at the leftmost edge of the simulation box  is collected, and we take a linear fitting of $log(a_{1L} / a_{1seed})$, where $a_{1seed} = 1\times 10^{-4}$. The growth rate of absolute SRS in Fig.~\ref{three_wave_2}(a) is $\gamma_{abs} = 0.07545 \gamma_{0}$. Then, with a stationary modulational wave number, $k_{s} = 0.03k_{0}$, the $\kappa_{m}$ dependence of absolute SRS growth rate  is shown in Fig.~\ref{three_wave_2}(b). Based on our three-wave simulations, the threshold of absolute SRS for $k_{s} = 0.03k_{0}$ is $\kappa_{m} \geq 0.06 \omega_{0} / c$  ($\varepsilon \geq 0.0108 $).  The growth rate becomes maximum  when $\kappa_{m} = 0.14 \omega_{0} / c$ ($ \varepsilon = 0.0252$), and the maximum growth rate is $\gamma_{abs} = 0.11979 \gamma_{0}$. Also, the wave number dependence of growth rate of absolute SRS with a fixed amplitude $\kappa_{m} = 0.1 \omega_{0} / c$  is shown in Fig.~\ref{three_wave_2}(c). The threshold of absolute SRS is $k_{s} \geq 0.02k_{0}$ when $\kappa_{m} = 0.1 \omega_{0} / c$. The peak of growth rate is also shown at around $k_{s} = 0.04k_{0}$ and the maximum growth rate is $\gamma_{abs} = 0.1054 \gamma_{0}$.

 The peak points of growth rate appear both in Fig.~\ref{three_wave_2}(b) and Fig.~\ref{three_wave_2}(c). The reason is that the derivative of wave number mismatch vanishes at the growth rate peak point, $i.e.$, $d\kappa / dx = \kappa' + k_{s}\kappa_{m} cos(k_{s} x) = 0$.\cite{sin} The vanishing condition is $\kappa' \sim k_{s}\kappa_{m}$. In Fig.~\ref{three_wave_2}(b), at the peak point, $k_{s}\kappa_{m} = 0.0042$, and in Fig.~\ref{three_wave_2}(c), $k_{s}\kappa_{m} = 0.004$ at the peak point. And $\kappa' \approx 0.0031$  in our case, which is close to the product of $k_{s}$ and $\kappa_{m}$ at the  peak points of the growth rate. Thus we understand the reason of existence of the growth rate peak points.

 Furthermore, as shown in Fig.~\ref{three_wave_2}(d), the red line is the threshold of absolute SRS at two dimensional plane of $k_{s}$ and $\kappa_{m}$ by using  Eq.~(\ref{wkb4}), and the grey shaded area represents the region where the absolute SRS can be induced by sinusoidal density modulation.  The blue diamonds in Fig.~\ref{three_wave_2}(d) are the threshold obtained by three-wave simulations, which agrees with Picard's quadratic fit formula in a specific wave number region ($k_{s} \leq 0.145 k_{0}$), The large wave number region will be discussed below.

\begin{figure}[htbp]
    \begin{center}
      \includegraphics[width=0.45\textwidth,clip,angle=0]{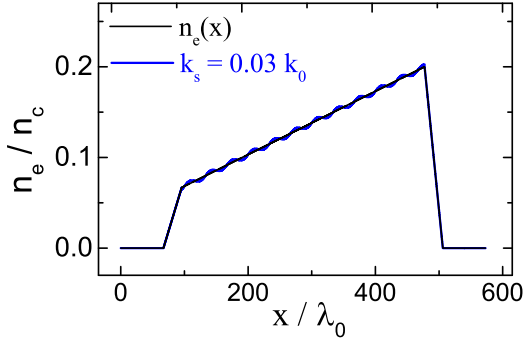}\vspace{-5pt}
      \caption{\label{myne} Diagram of plasma profile without density modulation (black line) and with density modulation $k_{s} = 0.03 k_{0}$, $ \varepsilon = 0.03 $(blue line).}
    \end{center}
  \end{figure}

  Then we consider the second situation and find that the system with short-wavelength density modulation ($k_{s} = 2 k_{0}$, $\kappa_{m} = 0.1 \omega_{0}/c$) cannot cause the absolute SRS. As shown in Fig.~\ref{three_wave}(b), the reflectivity (blue line) remains unchanged compared to that without density modulation (black line) in Fig.~\ref{three_wave}(a) and that obtained by Rosenbluth gain represented by green dashed line.  In Fig.~\ref{three_wave_2}(d), a sharp cut-off of short-wavelength modulation is plotted through three-wave simulations. The cut-off is seen as $k_{s} \leq 0.145 k_{0}$, equally, $\lambda_{s} = 2 \pi/k_{s} \geq 1.3 L_{0} $, which also appears in Picard's exact solution\cite{picard2} and Li's three wave simulations\cite{lij}. This result is also consistent with the requirement for absolute instability with imperfect laser and plasma conditions studied by many authors.\cite{laval,picard1,sin,turbulent,white,reiman} Because the absolute SRS need a basic length $L_{0}$ to grow, so the modulational length need to be comparable with $L_{0}$ to induce absolute SRS.

We conclude the three-wave simulations above, the absolute SRS can be induced by density modulation when $L_{m}\sim L_{0}$, and the threshold we obtained by three-wave simulations agrees with theoretical formula.\cite{picard2} But short-wavelength density modulation cannot induce absolute instability. We give the applicable range of wave number for Eq.~(\ref{wkb4}), $\lambda_{s} \geq 1.3 L_{0}$. However, the three-wave simulations can only consider the linear stage of SRS. Next, we would use the Vlasov simulations to study the influence of sinusoidal density modulation on SRS in inhomogeneous plasma at linear and nonlinear stage.

\section{Vlasov simulations }\label{Simulation model}

  A fully kinetic Vlasov code\cite{feng,wangqing2} is used here to study the SRS with density modulation. The plasma density profile is shown in Fig.~\ref{myne}. The density region is ranging  from $0.0667 n_{c}$ to $0.2 n_{c}$ with the formula $n_{e}(x) = n_{e0}[1+(x-x_{r})/L]$ (black line). And the blue line represents the plasma density with a sinusoidal density modulation, $n_{e}(x) = n_{e0}[1+(x-x_{r})/L + \varepsilon sin(k_{s}x)]$, where $L = 100 \mu m $, $\varepsilon = 0.03$, and $k_{s} = 0.03k_{0}$. The temperature of electron is $T_{e} = 1keV$. Ions are set to be immobile in Vlasov simulation to exclude SBS, because we only study SRS in Vlasov simulations.

  \begin{figure}
    \begin{center}
      \includegraphics[width=0.47\textwidth,clip,angle=0]{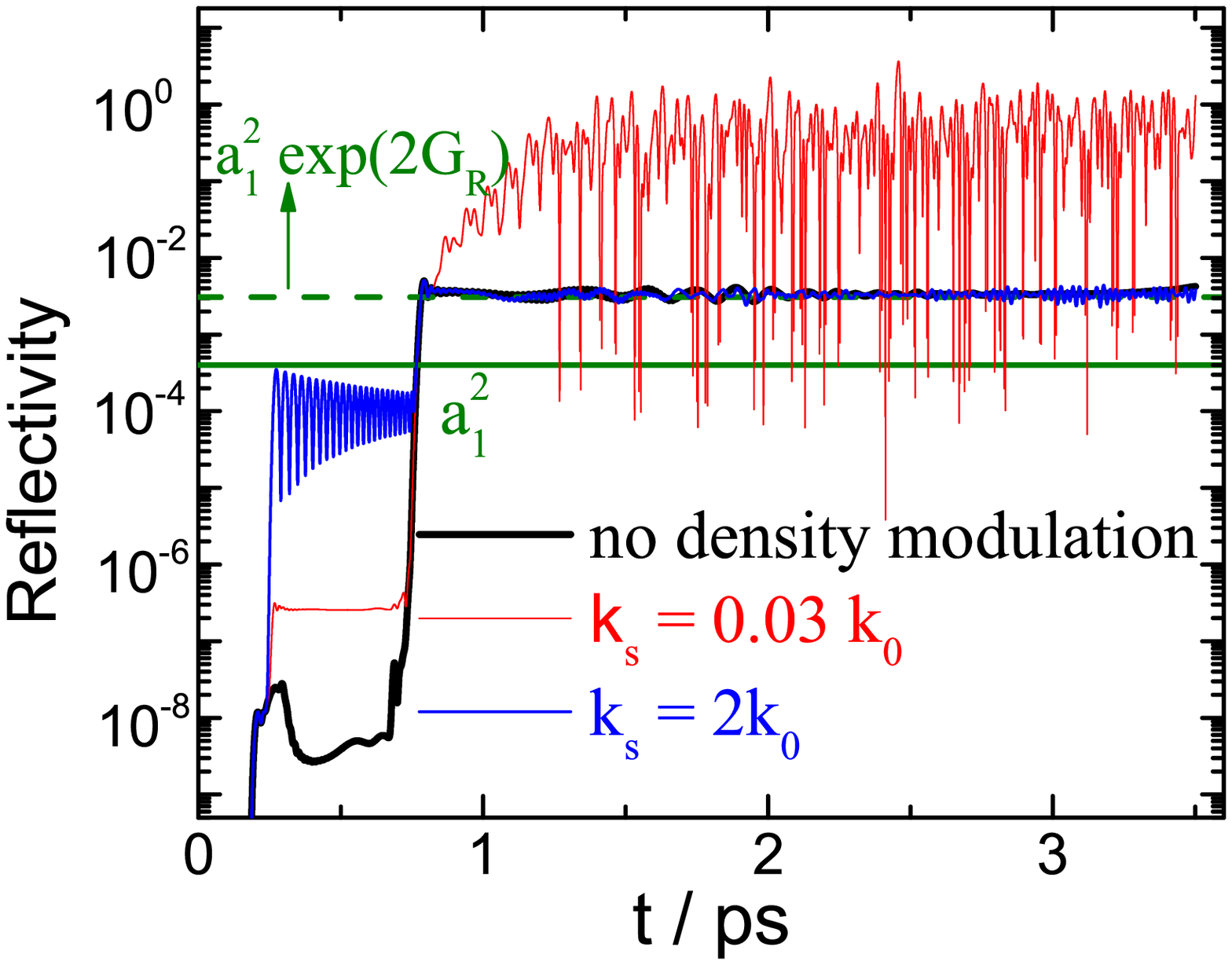}\vspace{-5pt}
      \caption{\label{vlasov_2}  The time history of reflectivity with $I_{0} = 4\times10^{15}W/cm^{2}$ without density modulation(black line),  red line represents the reflectivity with density modulation $k_{s} = 0.03 k_{0}$, $\varepsilon = 0.03$ and the blue line represents the reflectivity with density modulation  $k_{s} = 2 k_{0}$, $\varepsilon = 0.03$.}
    \end{center}
\end{figure}

 The whole simulation box is $L_{box} = 573 \lambda_{0}$ with a small length of vacuum space at the both sides of the simulation box, and we add a small length region with strong collision damping at both side of simulation box to avoid particle reflecting. And the space and time are discretized with $dx = 0.1 c / \omega_{0}$ and $dt = 0.1 \omega_{0}^{-1}$. The total simulation time is $t_{total} = 20000 \omega_{0}^{-1} = 3.5ps $. The pump laser enters into the simulation box at the left hand side with $\lambda_{0} = 0.351 \mu$m and intensity $I_{pump} = 1\times 10^{15} W/cm^{2} - 10\times 10^{15} W/cm^{2}$. The normalized light-intensity of pump laser is $a_{0} = 0.0095-0.03$. The seed laser enters into the plasma from the right side of the simulation box, The intensity of seed laser $a_{1} = 0.02 a_{0}$. And the wave length of is $\lambda_{1} = 0.530 \mu$m, which is corresponding to the SRS reflected light at $n_{e0} = 0.1 n_{c}$.

\subsection{ Linear stage: absolute SRS}\label{linear case}

 In the three-wave simulations, absolute SRS can be induced when the amplitude of density modulation is above the threshold.  The absolute SRS threshold is $\kappa_{m} = 0.06 \omega_{0} / c$ ($\varepsilon = 0.0108$) when the wave number of density modulation is  $k_{s} = 0.03k_{0}$. In our Vlasov simulation, the wave number and amplitude of long-wavelength case are $k_{s} = 0.03 k_{0}$ and $\varepsilon = 0.03 $, which is above the threshold of absolute SRS. Therefore, as shown in  Fig.~\ref{vlasov_2}, the red line shows a clear  feature of absolute SRS at $t = 0.8 ps \sim 1.2 ps$. The reflectivity is much larger than that without density modulation. We note that the reflectivity of SRS without density modulation is well agreed with the convective reflectivity represented by green dashed line using Eq.~(\ref{wkb3}). Then at $t > 1.2 ps$, the absolute SRS in Fig.~\ref{vlasov_2} comes into saturation because of nonlinear frequency shift. We will discuss this nonlinear phenomenon at the next subsection.

\begin{figure}[htbp]
    \begin{center}
      \includegraphics[width=0.44\textwidth,clip,angle=0]{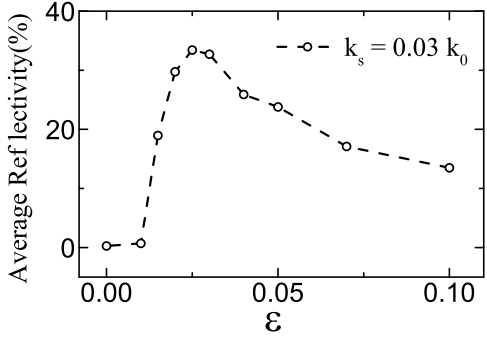}\vspace{-10pt}
      \caption{\label{vlasov_3} The density modulation amplitude $\varepsilon$ dependence of SRS reflectivity in an inhomogeneous plasma with long wave length modulation $k_{s} = 0.03 k_{0}$, and the intensity of pump laser is $I_{0} = 4\times10^{15}W/cm^{2}$. }
    \end{center}
  \end{figure}

Also, in Fig.~\ref{vlasov_3}, the black circles show the average reflectivity of SRS with varied amplitude of density modulation when the modulational wave number equals $0.03k_{0}$, the average reflectivity is obtained by $\bar{R} = R(t)/t_{total}$. When the amplitude of density modulation is $ \varepsilon = 0.01$, which is lower than the absolute SRS threshold, we observe that the reflectivity is still at a convective level. However, when $\varepsilon = 0.02$, which is above the threshold of absolute SRS, the average reflectivity increases to $29.75 \%$. As we know, the average reflectivity of absolute SRS is related to the temporal growth rate of SRS, thus, the growth rate of SRS in Fig.~\ref{three_wave_2} (b) has the same trend with the average reflectivity in Fig.~\ref{vlasov_3}. They both have a peak point in terms of a varying $\varepsilon$, and the peak point of growth rate located at $\varepsilon = 0.0252$ is consistent with that of average reflectivity located at $\varepsilon = 0.025$. Besides, from three-wave simulations, we know that the absolute SRS can only be induced when $\lambda_{s} \geq 1.3 L_{0}$. The blue line in Fig.~\ref{vlasov_2} is the time history of reflectivity when the short-wavelength density modulation ($k_{s} = 2k_{0}, \varepsilon = 0.03$) is added into the system.  The blue line is at the level of convective SRS reflectivity, because in this case $\lambda_{s} = 0.094 L_{0} < 1.3 L_{0}$, which is not in the region where absolute SRS can occur.

\subsection{ Nonlinear stage: suppression of inflation  }\label{inflation}

\begin{figure}
    \begin{center}
      \includegraphics[width=0.47\textwidth,clip,angle=0]{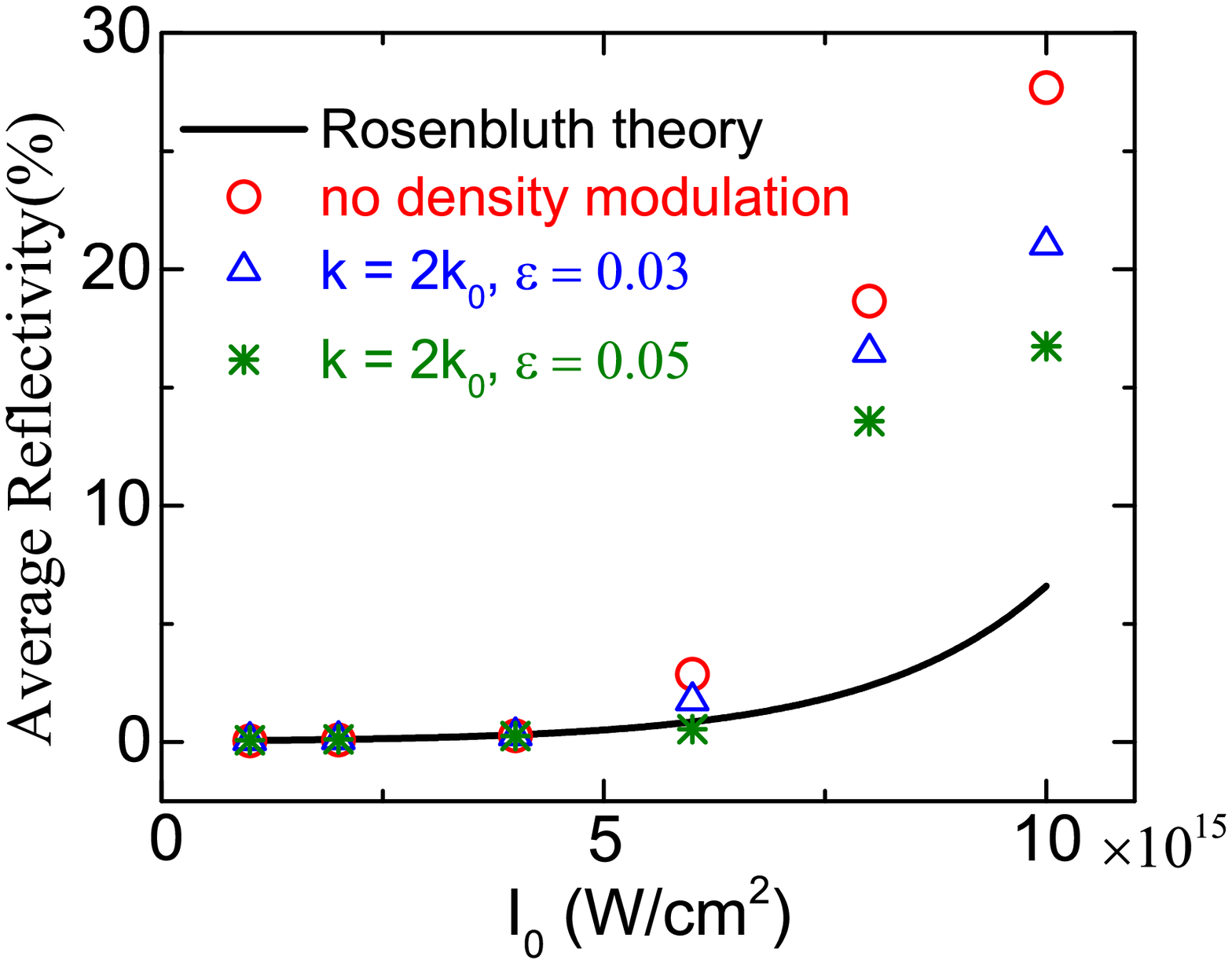}\vspace{-5pt}
      \caption{\label{vlasov_4}  The time history of reflectivity with $a_{1} = 0.08 a_{0}$ without density modulation(black line),  red line represents the reflectivity with density modulation $k_{s} = 0.03 k_{0}$, $\varepsilon = 0.03$ and the blue line represents the reflectivity with density modulation  $k_{s} = 2 k_{0}$, $\varepsilon = 0.03$.}
    \end{center}
\end{figure}

 As we know,  most nonlinear processes in SRS are related to the amplitude of Langmuir wave. When the amplitude of Langmuir wave is large enough to trap an amount of electrons near the phase velocity, the distribution function near phase velocity will be flattened, and then the Landau damping of Langmuir wave will be reduced accompanying  with the nonlinear frequency shift. Vu $et$  $al.$ reported that the burst of SRS reflectivity is induced by particle trapping, they call it inflation of SRS.\cite{hxvu1,hxvu2,hxvu3} And the nonlinear frequency shift will also happen, which is a saturation mechanism of SRS. The inflation and nonlinear frequency shift will cause pulse-like bursts of SRS. Similarly, in our Vlasov simulations, we find that the higher intensity of pump laser can result in the nonlinear stage of SRS.

 In Fig.~\ref{vlasov_4}, the black line is the pump laser intensity dependent theoretical reflectivity  based on the Rosenbluth gain. And the red circles represent the average reflectivity without density modulation. We observe that the simulations agree with Rosenbluth theory  when the intensity of pump laser is lower than $4\times 10^{15} W/cm^{2}$. This is because the SRS is in linear stage. When the intensity of pump laser is higher than $4\times 10^{15} W/cm^{2}$, the reflectivity is much higher than that obtained by Eq.~(\ref{wkb3}), which means nonlinear processes occur.

\begin{figure}[htbp]
    \begin{center}
      \includegraphics[width=0.47\textwidth,clip,angle=0]{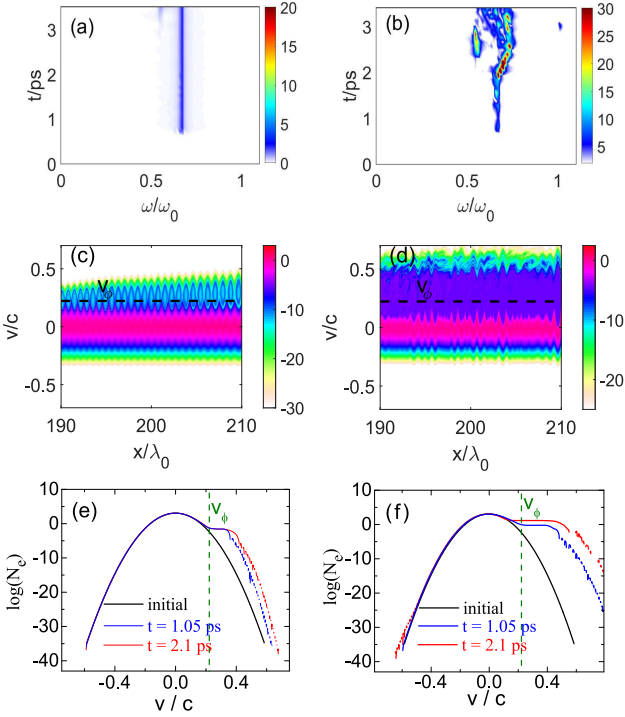}
      \caption{\label{vlasov_5} (a) the time history of frequency spectrum of reflected light with $I_{0} = 4\times10^{15} W/cm^{2}$ without density modulation. (b) the time history of frequency spectrum of reflected light with $I_{0} = 8\times10^{15} W/cm^{2}$ without density modulation. (c) The phase distribution of electron at $t = 2.1 ps$ with $I_{0} = 4\times10^{15} W/cm^{2}$. (d) The phase distribution of electron at $t = 2.1 ps$ with $I_{0} = 8\times10^{15} W/cm^{2}$. (e) The electron distribution function with $I_{0} = 4\times10^{15} W/cm^{2}$. (f) The electron distribution function with $I_{0} = 8\times10^{15} W/cm^{2}$.}
    \end{center}
  \end{figure}

 And we observe that the reflectivity has bursts during $2ps$ shown in Fig.~\ref{vlasov_5}(b), which is the frequency spectrum of reflected light when $I_{0}=8\times 10^{15} W/cm^{2}$. The burst of reflected light is not observed in Fig.~\ref{vlasov_5}(a) when the intensity of pump laser is $I_{0}=4\times 10^{15} W/cm^{2}$.

 Fig.~\ref{vlasov_5}(c) and Fig.~\ref{vlasov_5}(d) are the phase structure of electron near the phase velocity at $t = 2.1 ps$ for $I_{0}=4\times 10^{15} W/cm^{2}$ and $I_{0}=8\times 10^{15} W/cm^{2}$, respectively. The velocity width of trapping structure in Fig.~\ref{vlasov_5}(d) is larger than that in Fig.~\ref{vlasov_5}(c), and the trapping structure begin to merge when $I_{0}=8\times 10^{15} W/cm^{2}$. Correspondingly, the electron distribution functions are shown in  Fig.~\ref{vlasov_5}(e) and Fig.~\ref{vlasov_5}(f).  The electron distribution of $I_{0}=4\times 10^{15} W/cm^{2}$ keeps unchanged near the phase velocity from $t = 1.5ps$ to $t = 2.1 ps$, while the electron distribution in $I_{0}=8\times 10^{15} W/cm^{2}$ case is more flat and the width is much larger  near the phase velocity as time goes on. Based on the result of Ref.~[\onlinecite{hxvu1}],  particle trapping would reduce the Landau damping of Langmuir wave, which would cause the inflation of SRS. thus, we believe that the inflation of SRS happens when $I_{0}=8\times 10^{15} W/cm^{2}$. Also, we have changed the intensity of pump laser from $1\times 10^{15} W/cm^{2}$ to $10\times 10^{15} W/cm^{2}$, and we find that the  average reflectivity  is at the convective level  when $I_{0}=1\times 10^{15} W/cm^{2}$ to $I_{0}=4\times 10^{15} W/cm^{2}$, and increases to $18.66\%$ when $I_{0}=8\times 10^{15} W/cm^{2}$, This is the feature of inflation reported by Vu. \cite{hxvu1,hxvu2,hxvu3} Next, we will discuss the influence of sinusoidal density modulation on SRS at nonlinear stage.

\begin{figure}[htbp]
    \begin{center}
      \includegraphics[width=0.45\textwidth,clip,angle=0]{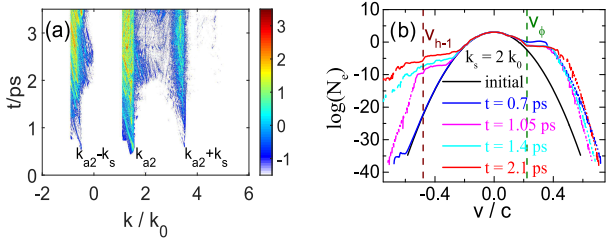}\vspace{-10pt}
      \caption{\label{vlasov_6} (a) The time history of wave number spectrum of Langmuir wave with $I_{0} = 8\times10^{15} W/cm^{2}$ and density modulation ($k_{s} = 2 k_{0}$, $\varepsilon = 0.05$). (b) The distribution function of electron with $I_{0} = 8\times10^{15} W/cm^{2}$ and density modulation ($k_{s} = 2 k_{0}$, $\varepsilon = 0.05$).}
    \end{center}
  \end{figure}


At last, we study the influence of short-wavelength density modulation on the SRS at nonlinear stage. As shown in the blue triangles and green stars of Fig.~\ref{vlasov_4}, the reflectivity  keeps  unchanged as the the amplitude of short-wavelength density modulation when $I_{0} <= 4 \times 10^{15} W/cm^{2}$. This agrees  well with our three-wave simulations.  And the SRS reflectivity decreases with modulational amplitude when $I_{0} >= 6 \times 10^{15} W/cm^{2}$. In Ref.~[\onlinecite{Barr,Curtet}], the Landau damping of Langmuir wave increases with the modulational amplitude. This is one of the reason that SRS reflectivity decreases with modulational amplitude at nonlinear stage.

And O'Neil \cite{Neil} discussed that the particle trapping can reduce the Landau damping for $\omega_{bounce} t > 2\pi$, where $\omega_{bounce} = \omega_{pe} k\lambda_{De}\sqrt{\frac{e\phi}{T_{e}}}$, $\sqrt{\frac{e\phi}{T_{e}}}$ is the amplitude of Langmuir wave. So, reducing the amplitude of Langmuir wave is an effective way to avoid inflation of SRS.

When we add the short-wavelength modulation, the harmonic waves will be induced. Fig.~\ref{vlasov_6}(a) is the time history of plasma wave in $k$ space, and we can observe two clear harmonic waves \cite{haochu} at $k_{a_{2}}-k_{s}$ and $k_{a_{2}}+k_{s}$, where $k_{s} = 2 k_{0}$. The wave number of primary Langmuir wave is $k_{a_{2}} = 1.53 k_{0}$, so the wave number of the downward harmonic wave is $k_{h-1} = -0.47 k_{0} $, which means the propagation direction of downward harmonic wave is opposite to the Langmuir wave. The downward harmonic wave has higher intensity than the upward harmonic wave, because the downward harmonic wave has a lower Landau damping. The downward harmonic wave can take energy away from Langmuir wave and then suppress SRS.
As depicted in  Fig.~\ref{vlasov_6}(b), the green dashed line represents the phase velocity of Langmuir wave, $v_{\phi} = 0.22 c$, corresponding to the velocity of electron trapping, and the brown dash line is the phase velocity of downward harmonic wave.  In theory, the phase velocity should be $-0.66c$, but in simulation, $v_{h-1} = -0.48 c$  because of the broadening of downward harmonic wave in $k$ spectrum as shown in Fig.~\ref{vlasov_6}(a). The $k$ spectrum of Langmuir wave broadens to around $1.3 k_{0}$, so the real trapping velocity of downward harmonic wave is at around $ -0.48 c$ as shown in  the electron distribution function at different times. The electron trapping around primary Langnuir wave velocity keeps in a same level because of the downward harmonic wave, and more electrons are trapped by downward harmonic wave near $v_{h-1}$ from $ t \sim 1.05 ps$ to $t \sim 2.1 ps$. We notice that there is no electron trapping in the minus $v$ direction in the case of no density modulation in Fig.~\ref{vlasov_5}(f). So, the suppression of the inflation is the consequence of  downward harmonic wave induced by short-wavelength density modulation.

In Fig.~\ref{vlasov_4}, the average reflectivity of SRS decreases with the amplitude of density modulation at nonlinear stage. This phenomenon is different from the long-wavelength modualtional case showed in Fig.~\ref{vlasov_3}. The long-wavelength density modulation can easily cause the absolute SRS to increase the reflectivity of SRS. Thus, in ICF experiments, one should try to avoid the long-wavelength density modulation, and one would add the short-wavelength density modulation to suppress the inflation of SRS.

\section{PIC simulations }\label{PIC model}
\begin{figure}
    \begin{minipage}[t]{0.5\linewidth}
    \centering
    \includegraphics[width=1.96in]{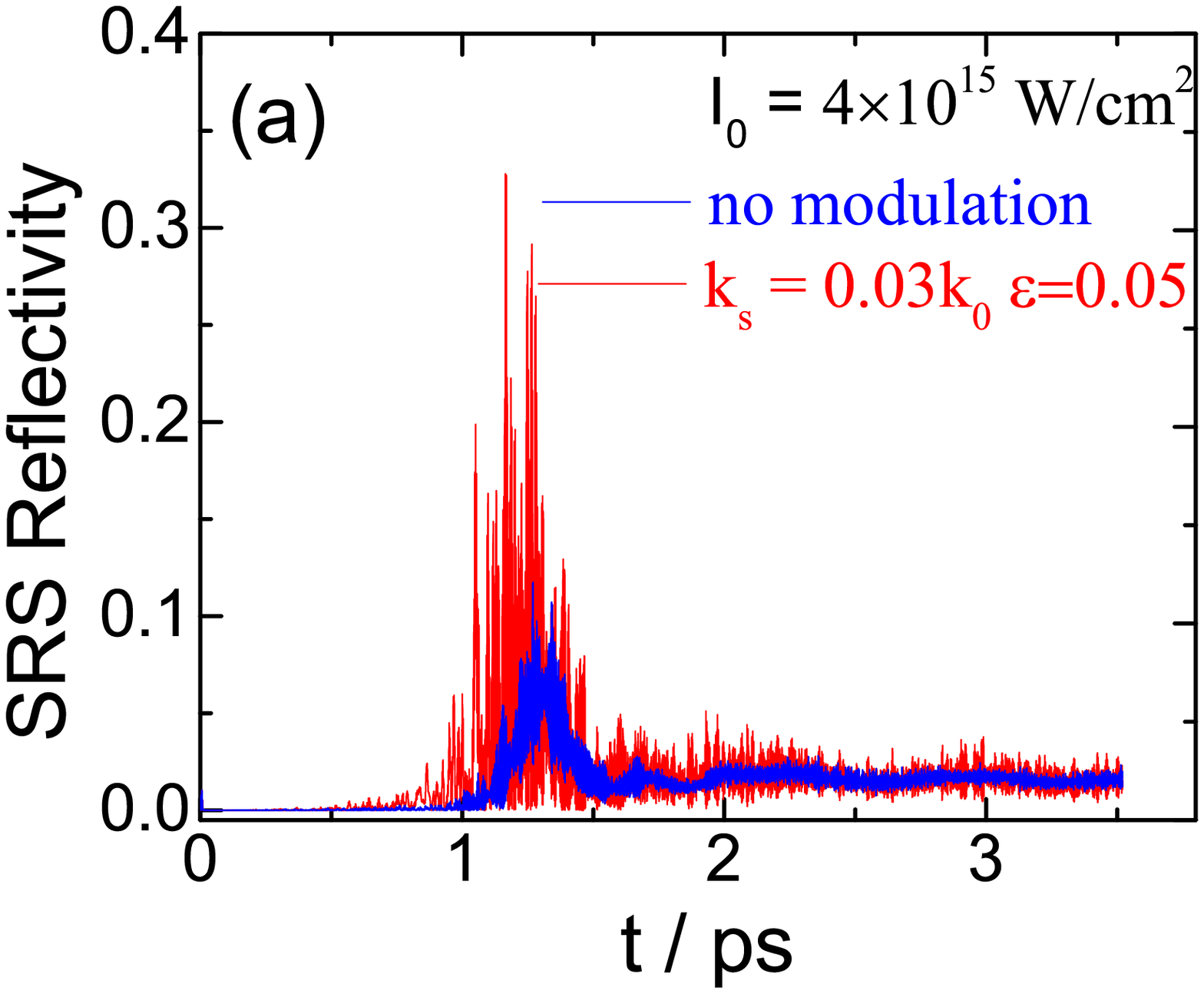}\vspace{-18pt}
    \label{fig:side:a}
    \end{minipage}%
    \begin{minipage}[t]{0.5\linewidth}
    \centering
    \includegraphics[width=1.96in]{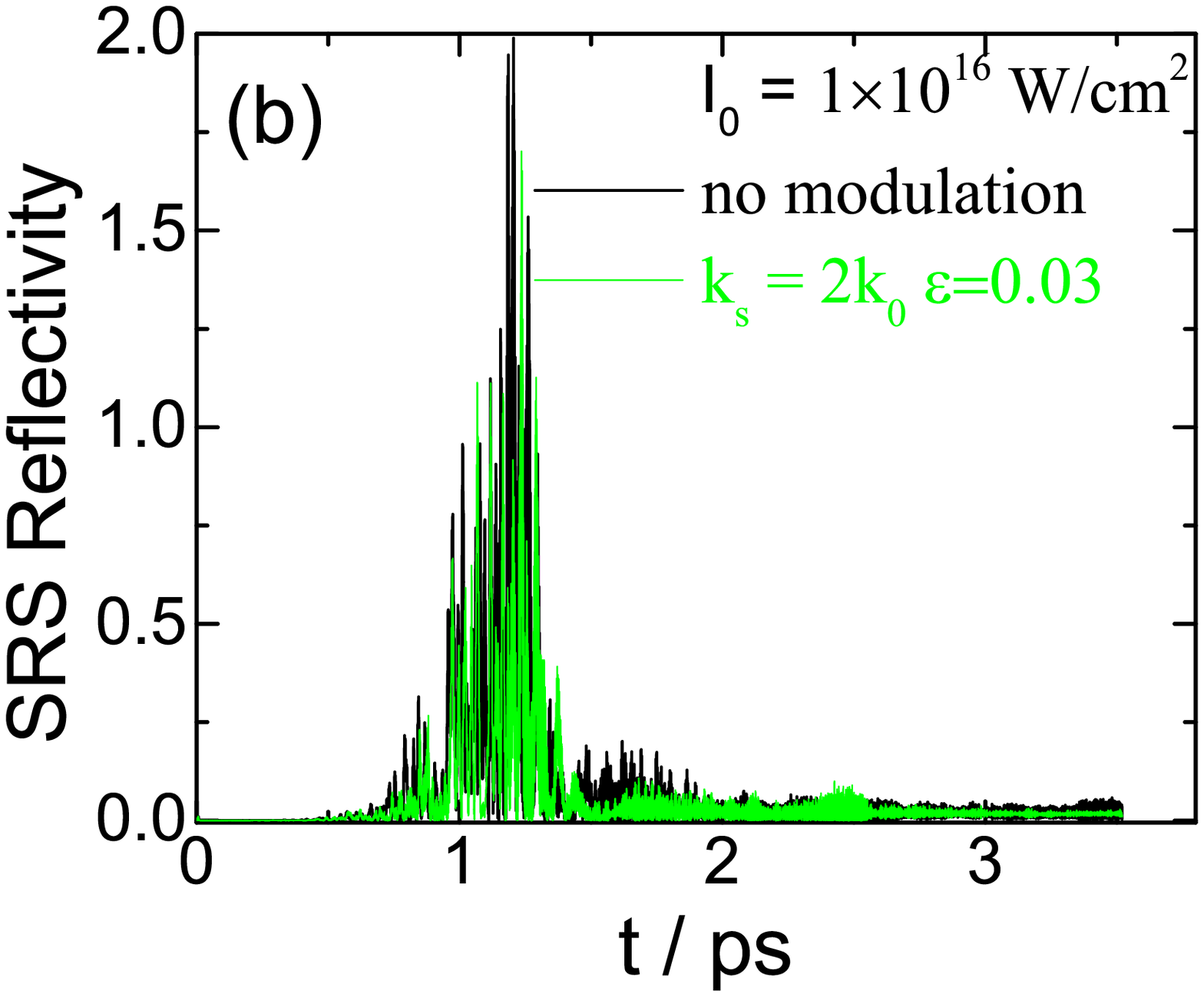}\vspace{-18pt}
    \label{fig:side:b}
    \end{minipage}
    \caption{\label{pic_1}(a) The time history of SRS reflectivity without density modulation(blue line), and that with long-wave length density modulation(red line). The intensity of pump laser is $I_0=4\times10^{15}W/cm^{2}$.  (b) The time history of SRS reflectivity without density modulation(black line), and that with short-wave length density modulation(green line). The intensity of pump laser is $I_0=1\times10^{16}W/cm^{2}$. }
\end{figure}

In our three-wave simulations and Vlasov simulations, we choose a specific plasma density $n_{e} =0.1 n_{c}$ as the resonant density.  And in our Vlasov simulations, we added a seed laser and did not consider the motion of ions. All these make our model too limited. In order to test our model in a more general situation, now we use PIC code EPOCH\cite{chen}.  In our PIC simulations, the electron density profile is same as our Vlasov simulations, and we use $He^{2+} (M = 7344m_{e},Z = 2)$ as the ion. The temperature of electron is $T_{e} = 1keV$, and the temperature of ion is $T_{i} = 0.5keV$. And the seed laser is not considered in our PIC simulations.

\begin{figure}[htbp]
    \begin{center}
      \includegraphics[width=0.45\textwidth,clip,angle=0]{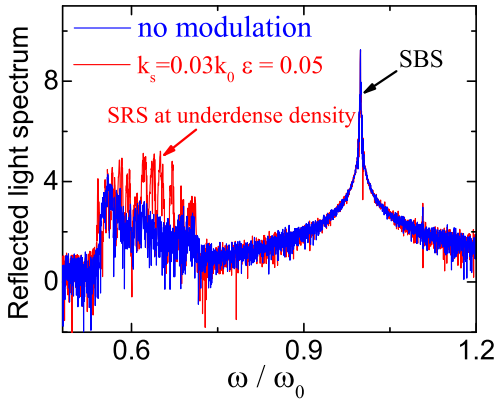}\vspace{-5pt}
      \caption{\label{pic_2} Reflected light spectrum  without  density modulation (blue line) and with density modulation $k_{s} = 0.03 k_{0}$, $ \varepsilon = 0.05 $ (red line). The intensity of pump laser is $I_0=4\times10^{15}W/cm^{2}$. }
    \end{center}
  \end{figure}

First, we study the influence of long-wavelength modulation on SRS at linear stage. As shown in Fig.~\ref{pic_1} (a), the intensity of pump laser is $I_{0} = 4\times10^{15} W/cm^{2}$,  and we filter out the SRS reflected light. The blue line is the SRS reflectivity without density modulation, and the red represents the SRS reflectivity when the long-wave length modulation is present. At early time, $t < 1.5 ps$, the SRS is dominant, and the SBS becomes main instability when $t > 1.5 ps$. We observe that the SRS reflectivity is more intense than that without density modulation. This is because the long-wavelength density modulation induce the SRS at lower plasma density. Fig.~\ref{pic_2} is the corresponding reflected light spectrum, the reflected light at $\omega = 0.6\omega_{0}-0.7\omega_{0}$ is improved by long-wavelength modulation. Because the long-wavelength modulation induces the absolute SRS at $n_{e} = 0.075n_{0}-0.15n_{c}$, which is agreement with our three wave simulations and Vlasov simulations. Next, we study the influence of short-wavelength modulation on SRS at nonlinear stage. In Fig.~\ref{pic_1} (b), we filter the SRS reflected light when $I_{0} = 1\times10^{16} W/cm^{2}$. The black line is the SRS reflectivity without density modulation, and the green line represents the SRS reflectivity with short-wavelength modulation $k_{s} = 2k_{0}, \epsilon = 0.03$. We observe that the SRS reflectivity decreases when short-wavelength is present, the max value changes from 2 to 1.6, and the average SRS reflectivity changes from $7\%$ to $5\%$. This phenomenon is qualitatively agreement with our Vlasov simulations.

\section{Conclusion and discussion}\label{conclusion}

 In this paper, at first, we numerically solve the three-wave equations to study the influence of density modulation on the SRS reflectivity in  underdense  inhomogeneous plasma. We find that the long-wavelength density modulation is able to destabilize the convective SRS and cause a absolute SRS when the modulational amplitude is above the threshold. And the absolute SRS threshold has a sharp cut off, $\lambda_{s} \geq 1.3 L_{0}$, which agrees with the exact solution of Picard and Johnston\cite{picard2}. Second, the Vlasov simulation is used to study the influence of density modulation at linear and nonlinear stage.
 At linear stage, the Vlasov simulations are well consistent with the numerical solution of three-wave equations. The average reflectivity has the same trend with the temporal growth rate. And the short-wavelength density modulation $i.e.$ $k_{s} = 2k_{0}$ can not induce absolute SRS. Then, we find that the inflation of SRS would be induced by particle trapping when $I_{0} \geq 6\times10^{15} W/cm^{2}$.  And the pulse-like burst of reflectivity is caused by inflation and nonlinear frequency shift. At last, we study the short-wavelength modulation $i.e.$ $k_{s} = 2 k_{0}$ influence on the inflation of SRS. The Landau damping increases with modulational amplitude. And the downward harmonic wave is an effective way to take away the energy of Langmuir wave, thus the reflectivity of SRS will decrease. Finally, our PIC simulations are qualitatively consistent with our Vlasov simulations.

In the early experiments, the lower threshold of SRS than Rosenbluth threshold in the underdense  plasma has been observed,\cite{Tanaka,Phillion1,Phillion2,Turner}  Estabrook and Kruer pointed out that these phenomena may be related to the plasma noise through simulations.\cite{Kent} In this paper, we offer an alternative explanation, that is the intense SRS reflected light in the low density plasma may be caused by the long wave length density modulation. Based on our simulations, the absolute SRS can be induced by a density modulation with very small amplitude $ \epsilon \sim 1 \%$. The sinusoidal density modulation we consider here are similar to the turbulent density gradient,\cite{turbulent} both these two kinds of modulation can induce absolute SRS. Besides, our results show that the short wave length density modulation has the ability to suppress the inflation of SRS by the downward harmonic wave. This mechanism may be applied in the high intensity and inhomogeneous ignition schemes\cite{D_D,H_D,S_I,Klimo1,Klimo2,HaoL1} to suppress SRS.

\section*{Acknowledgements}
We are pleased to acknowledge useful discussions with Q. Wang, Y. Z. Zhou and L. Hao.  This work was supported by the Strategic Priority Research Program of Chinese Academy of Sciences (Grant No. XDA25050700), National Natural Science Foundation of China (Grant Nos. 11805062, 11875091 and 11975059), Science Challenge Project, No. TZ2016005 and Natural Science Foundation of Hunan Province, China (Grant No. 2020JJ5029).


\begin{thebibliography}{100}
\bibitem{ICF1}R. Betti, C. D. Zhou, K. S. Anderson, L. J. Perkins, W. Theobald, and A. A. Solodov,  Phys. Rev. Lett. {\bf 98}, 155001 (2007).
\bibitem{ICF2}Basov N. G., Guskov S. Y. and Feokistov L. P., J. Russ. Laser Res. {\bf 13}, 396 (1992).
\bibitem{ICF3}Max Tabak, James Hammer, Michael E. Glinsky, William L. Kruer, Scott C. Wilks, John Woodworth, E. Michael Campbell, and Michael D. Perry,  Phys. Plasmas {\bf 1}, 1626 (1994).
\bibitem{xiao2}C. Z. Xiao, H. B. Zhuo, Y. Yin, Z. J. Liu, C. Y. Zheng, Y. Zhao and X. T. He,  Plasma Phys. Control. Fusion {\bf 60},025020 (2018).
\bibitem{xiao3} C. Z. Xiao, H. B. Zhuo, Y. Yin, Z. J. Liu, C. Y. Zheng, and X. T. He, Phys. Plasmas {\bf 26}, 062109 (2019).
\bibitem{xiao4} C. Z. Xiao,  H. B. Zhuo, Y. Yin, Z.J. Liu, C.Y. Zheng, and X.T. He, Nucl. Fusion {\bf 60}, 016022 (2020).
\bibitem{kruer}W. L. Kruer, \emph{The Physics of Laser Plasma Interactions} (Westview Press, Boulder, 2003).
\bibitem{Nicholson} D. R. Nicholson, \emph{Introduction to Plasma Theory }(John Wiley \& Sons, New York, 1983).
\bibitem{xiao1} C. Z. Xiao, Z. J. Liu, D. Wu, C. Y. Zheng, and X. T. He Phys. Plasmas {\bf 22}, 052121 (2015).
\bibitem{yanxia} Y. X. Wang, Q. S. Feng, H. C. Zhang, Q. Wang, C. Y. Zheng, Z. J. Liu, and X. T. He, Phys. Plasmas {\bf 24}, 103122 (2017).
\bibitem{haochu} H. C. Zhang, C. Z. Xiao, Q. Wang, Q. S. Feng, Z. J. Liu, and C. Y. Zheng, Phys. Plasmas {\bf 24}, 032118 (2017).

\bibitem{chen} Y. Chen, C. Y. Zheng, Z. J. Liu, L. H. Cao, Q. S. Feng and C. Z. Xiao, Plasma Phys. Control. Fusion {\bf 62}, 105020 (2020)

\bibitem{I_D}   J. Lindl, Phys. Plasmas {\bf 2}, 3933 (1995).

\bibitem{D_D}  S. E. Bodner, D. G. Colombant, J. H. Gardner, R. H. Lehmberg, S. P. Obenschain, L. Phillips, A. J. Schmitt, J. D. Sethian, R. L. McCrory, W. Seka, C. P. Verdon, J. P. Knauer, B. B. Afeyan, and H. T. Powell, Phys. Plasmas {\bf 5}, 1901 (1998).
\bibitem{H_D} X. T. He, J.W. Li, Z. F. Fan, L. F.Wang, J. Liu, K. Lan, J. F.Wu, and W. H. Ye, Phys. Plasmas {\bf 23}, 082706 (2016).

\bibitem{S_I} X. Ribeyre, G. Schurtz, M. Lafon, S. Galera, and S. Weber, Plasma Phys. Control. Fusion {\bf 51}, 015013 (2009).

\bibitem{rosenbluth1} M. N. Rosenbluth, Phys. Rev. Lett. {\bf 29}, 565 (1972).
\bibitem{rosenbluth2} M. N. Rosenbluth, R. B. White, C. S. Liu, Phys. Rev. Lett. {\bf 31}, 1190 (1973).





\bibitem{laval} G. Laval, R. Pellat and D. Pesme, Phys. Rev. Lett. {\bf 36}, 192 (1974).
\bibitem{picard1} G. Picard and T. W. Johnston, Phys. Rev. Lett. {\bf 54}, 574 (1983).
\bibitem{lij} J. Li, R. Yan, and C. Ren, Phys. Plasmas {\bf 24}, 052705 (2017).
\bibitem{picard2} G. Picard and T. W. Johnston, Phys. Fluids {\bf 28}, 859 (1985).
\bibitem{sin} D. R. Nicholson Physics of Fluids  {\bf 19}, 889 (1976).
\bibitem{turbulent} D. R. Nicholson and A. N. Kaufman, Phys. Rev. Lett. {\bf 33}, 1207 (1974).
\bibitem{white} R. White, P. Kaw, D. Pesme, M. N. Rosenbluth, G. Laval, R. Huff, R. Varma,  Nucl. Fusion {\bf 14} 45 (1974)
\bibitem{reiman}  A. H. Reiman, A. Bers, and D. J. Kaup, Phys. Rev. Lett. {\bf 39}, 850 (1977)
\bibitem{feng} Q. S. Feng, C. Y. Zheng, Z. J. Liu, L. H. Cao, Q. Wang, C. Z. Xiao, and X. T. He, Phys. Plasmas {\bf 25}, 092112 (2018)
\bibitem{wangqing2} Q. Wang, C. Y. Zheng, Z. J. Liu, L. H. Cao, Q. S. Feng, C. S. Liu and X. T. He, Plasma Phys. Control. Fusion {\bf 61}, 085017 (2019)
\bibitem{yang}  S. J. Yang, Y. Chen, and C. Z. Xiao, AIP Advances {\bf 10}, 065208 (2020).
\bibitem{hxvu1} H. X. Vu, D. F. DuBois, and B. Bezzerides, Phys. Rev. Lett. {\bf 86}, 4306 (2001).
\bibitem{hxvu2} H. X. Vu, D. F. DuBois, and B. Bezzerides, Phys. Plasmas {\bf 9}, 1745 (2002).
\bibitem{hxvu3} H. X. Vu, D. F. DuBois, and B. Bezzerides, Phys. Plasmas {\bf 14}, 012702 (2007)

\bibitem{Klimo1} O. Klimo, S. Weber, V. T. Tikhonchuk and J. Limpouch, Plasma Phys. Control. Fusion {\bf 52},  055013 (2010).
\bibitem{Klimo2}  O. Klimo, V. T. Tikhonchuk, X. Ribeyre, G. Schurtz, C. Riconda, S. Weber, and J. Limpouch, Phys. Plasmas {\bf 18}, 082709 (2011).
\bibitem{HaoL1}   L. Hao, J. Li, W. D. Liu, R. Yan, and C. Ren, Physics of Plasmas {\bf 23}, 042702 (2016);

\bibitem{sweby}  Sweby P. K., J. Numer. Anal. {\bf 21}, 995 (1984)
\bibitem{wangqing1}  Q. Wang, C. Y. Zheng, Z. J. Liu, C. Z. Xiao, Q. S. Feng, H. C. Zhang and X. T. He, Plasma Phys. Control. Fusion {\bf 60}, 025016 (2018)
\bibitem{Tanaka} K. Tanaka, L. M. Goldman, W. Seka, M. C.Richardson, J. M. Soures, and E. A. Williams  Phys. Rev. Lett. {\bf 48}, 1179 (1982).
\bibitem{Phillion1} D. W. Phillion, E. M. Campbell, K. G. Estabrook, G. E. Phillips, and F.Ze, Phys. Rev. Lett. {\bf 49},1405 (1982).
\bibitem{Phillion2} D. W. Phillion, D. L. Banner, E. M. Campbell, R. E. Turner, and K. G. Estabrook, Phys. Fluids {\bf 25},1434 (1982).
\bibitem{Turner} R. E. Turner, D. W. Phillion, E. M. Campbell, and Kent Estabrook, Phys. Fluids {\bf 26},579 (1983).
\bibitem{Kent}   Kent Estabrook and W. L. Kruer,  Phys. Rev. Lett. {\bf 53},465 (1984)

\bibitem{Barr}   Hugh C. Barr, and Francis F. Chen, Phys. Fluids {\bf 30}, 1180 (1987)
\bibitem{Curtet} Curtet M. and Bonnaud G., Phys. Rev. E, {\bf 60}, 5 (1999)
\bibitem{Neil}   T. M. O'Neil, Phys. Fluids. {\bf 8}, 2255 (1965)

\end{thebibliography}

\end{document}